# Colloidal Synthesis of Strongly Fluorescent CsPbBr$_3$ Nanowires with Width Tunable Down to the Quantum Confinement Regime

Muhammad Imran,[†,&] Francesco Di Stasio,[†] Zhiya Dang,[†] Claudio Canale,[‡] Ali Hossain Khan,[†] Javad Shamsi,[†,&] Rosaria Brescia,[†] Mirko Prato[†]* and Liberato Manna[†]*

[†] Nanochemistry and [‡] Nanophysics Depts., Istituto Italiano di Tecnologia, Via Morego 30, 16163 Genova, Italy
[&] Dipartimento di Chimica e Chimica Industriale, Università degli Studi di Genova, Via Dodecaneso, 31, 16146, Genova, Italy

*Supporting Information Placeholder*

**ABSTRACT:** We report the colloidal synthesis of strongly fluorescent CsPbBr$_3$ perovskite nanowires (NWs) with rectangular section and with tuneable width, from 20 nm (exhibiting no quantum confinement, hence emitting in the green) down to around 3 nm (in the strong quantum-confinement regime, emitting in the blue), by introducing in the synthesis a short acid (octanoic acid or hexanoic acid) together with alkyl amines (octylamine and oleylamine). Temperatures below 70 °C promoted the formation of monodisperse, few unit cell thick NWs that were free from by-products. The photoluminescence quantum yield of the NW samples went from 12% for non-confined NWs emitting at 524 nm to a maximum of 77% for the 5 nm diameter NWs emitting at 497 nm, down to 30% for the thinnest NWs (diameter ~ 3nm), in the latter sample most likely due to aggregation occurring in solution.

Semiconductor nanowires[1] (NWs) have received considerable attention as candidates in a wide variety of applications, such as in optoelectronics,[2-4] photovoltaics,[5] thermoelectrics,[6] and sensing.[7] Several studies have addressed the fabrication of strongly quantum confined NWs, for example made of SnO$_2$[8] or cadmium chalcogenides.[9] The rapid emergence of lead halide perovskites as promising materials in photovoltaics and optoelectronics[10, 11] has recently placed NWs under the spotlight again. For example, NWs based on methyl ammonium lead halide perovskites (CH$_3$NH$_3$PbX$_3$), prepared by a surface-initiated solution growth,[12] were reported to have low lasing thresholds and high quality factors, PLQYs close to 100% and broad tunability through the whole visible range. Similar results have been obtained also with fully inorganic CsPbX$_3$ NWs.[13] Colloidal approaches have been proposed for both hybrid[14] and fully inorganic[15, 16] perovskite NWs. For hybrid lead halide perovskites, CH$_3$NH$_3$PbBr$_3$ NWs could be grown up to 900 nm in length and their width could be tuned by varying the reaction time, such that blue emitting NWs (hence strongly confined) were formed at short reaction times, and green emitting ones (not-confined) at longer times. Fully inorganic Cs-based perovskites are less susceptible to hydrolysis from moisture than their hybrid counterparts,[17] and therefore are preferable for applications. However, protocols developed to date could deliver wires of up to 5 microns length, but not with width in the strong quantum confinement regime.

Here, we report a colloidal synthesis of CsPbBr$_3$ NWs with width that is tunable down to the quantum-confined regime (few-unit-cell thick), using a mixture of alkyl amines and a short alkyl carboxylic acid as growth medium, following standard air-free techniques (for details see Figure S1 and Table S1 of the Supporting Information, SI). In our initial scheme (similar to that of Zhang et al.[16]), we could synthesize NWs with 10-20 nm width (hence non-confined) by regulating the ratio of octylamine (OctAm) to oleylamine (OlAm) and by varying the reaction time (from 30 to 50 min), see Figures S1a, S2 and S3. The addition of an acid with a long alkyl chain (oleic acid) combined with short and long chain amines led instead to the formation of nanosheets (Figure S4), as recently shown by us.[18] If, in *lieu* of oleic acid, a shorter alkyl chain carboxylic acid (octanoic acid–OctAc, or hexanoic acid–HexAc) was used, thinner NWs could be prepared (Figure S5). By increasing the concentration of the short acid over that of the amine ligands (OctAm and OlAm), the width of nanowires could be tuned from 10±1 to 3.4±0.5 nm, that is, down to the strong quantum confinement regime (Table S1 and Figure S6). We could even grow NWs that were around 2.8 nm thick (see Table S1 and Figure S6f), but they were too unstable over time and were not considered further. Other parameters that were found critical to control the width of the NWs were the temperature and the reaction time. For non-confined nanowires (width≥10nm) 120-130°C was the optimal temperature range for growth. For the growth of confined NWs (width ≤10nm) the same temperature range (or higher temperature) yielded various by-products (including cubes) in addition to wires, whereas below 70°C the NW growth was much slower, which helped to improve the size mono dispersity and led essentially to NWs free of byproducts (see Figure S7). The optimal time of growth was 50 minutes for both non-confined and confined NWs. Increasing the reaction time resulted in aggregation of the NWs.

Bright field transmission electron microscopy (BF-TEM) images are reported in Figure 1 (a-c), along with high-angle angular dark field scanning TEM (HAADF-STEM) images (d, f, g) and histograms of widths distributions (e, g, i), for NW samples prepared under different conditions (see Figure S6 and Table S1 for details). High resolution TEM (HRTEM) analyses revealed the single-crystal nature of the NWs, compatible with the orthorhombic structure (ICSD #97851) already reported by us for $CsPbBr_3$ nanosheets[18] and by Zhang et al.[15] for NWs. Due to the well-known electron-beam-irradiation sensitivity of lead halide perovskite NCs in general (both the hybrid[19] and the fully inorganic[20] ones), and to the extremely small size of the most confined NWs reported here, HRTEM analyses of these few-nm wide nanocrystals were challenging. For this reason, HRTEM images shown in the current work were collected using a direct-electron detection camera instead of the more conventional CCD camera, by which we could obtain, while using short acquisition times, high resolution images at relatively low electron dose and wide field of view (FOV). A careful look at these results revealed that the NWs do not have a circular but rather a rectangular section, as each sample was made of bundles of NWs exhibiting identical orientation (*i.e.*, zone axis), as can be seen in the wide FOV HRTEM images in Figure 2 (a,e,i) and their respective FFTs (panels b,f,j). In particular, all the NWs were enclosed by {110}, {1-10} and {001} facets and all of them were confined along the <110> direction. However, the 10 nm thick NWs on the carbon support film were <1-10>-oriented and elongated along the <001> direction, while thinner ones (both 5.1 nm and 3.4 nm) were <001>-oriented and elongated along the <1-10> direction. This difference can be ascribed to the addition of short chain acids in the synthesis, which clearly favor the growth along <1-10> while inhibiting the growth along the <001> direction.[15, 16] An additional indication of the rectangular section of the NWs, in the particular case of the 10 nm wide NW sample, comes from the two different values measured for the size along the transverse direction, hinting at width and thickness of the NWs (see Figure S8 and related discussion in the SI).

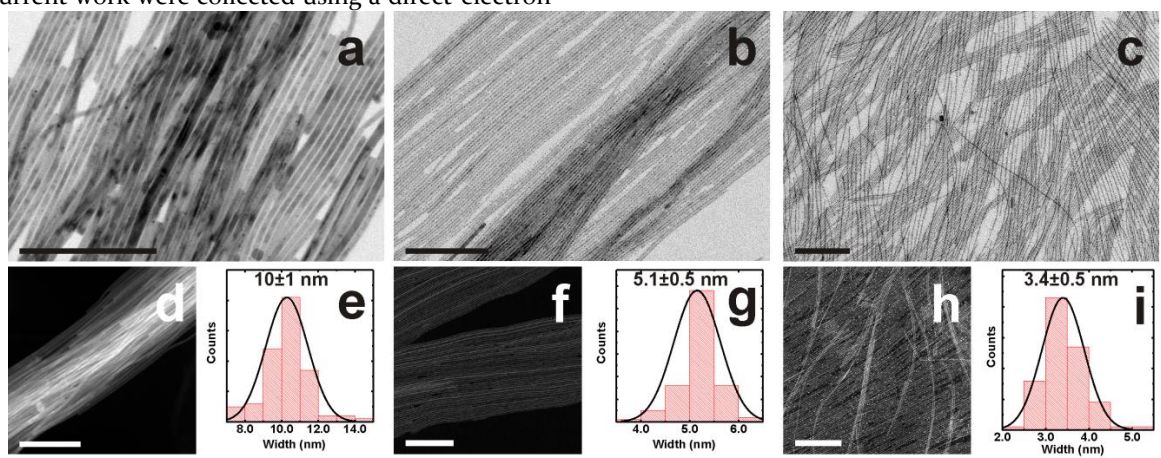

**Figure 1.** Effect of increasing the ratio of short chain carboxylic acid to amine ligands on controlling the width of the $CsPbBr_3$ NWs. Representative BF-TEM and HAADF-STEM images for 10 nm (a,d), 5.1 nm (b,e) and 3.4 nm (c,f) width NWs and (e,g,i) their respective size distributions. Short chain carboxylic acid to alkyl amines volume ratios respectively: 0 (a), 0.1 (b), 0.3 (c) (see Table S1). Scale bars are 200 nm in all images.

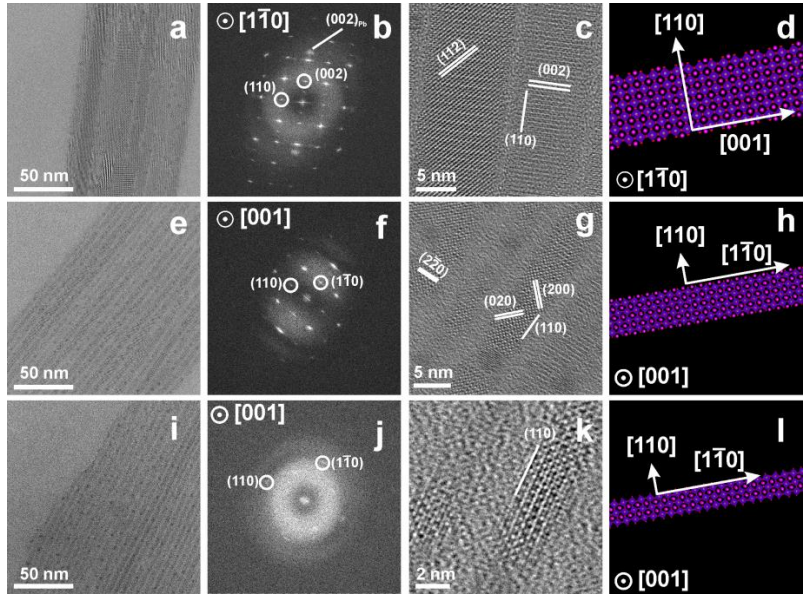

**Figure 2.** Structural analyses of (a-d) 10 nm, (e-h) 5.1 nm and (i-l) 3.4 nm-wide NWs: (a,e,i) wide FOV HRTEM images of the NWs, (b,f,j) corresponding FFTs and (c,g,k) magnified view of respective portions; (d,h,l) schematic crystallographic models of the NWs, showing the orientation of the observed facet and the elongation direction for the NWs.

We conclude that the synthesis conditions inhibited the growth of some crystallographic facets, while enhancing three sets of them, namely the {110}, {1-10} and {001} ones, and the relative stability of the facets was likely related to the concentration of the short chain acids used. Also, epitaxially oriented Pb nanocrystals were found along the NWs (see Figure 2(a,b,e,f,i)). Their formation was due to electron-beam irradiation (despite the low dose), as previously reported for cesium lead halide nanocrystals.[18, 20] In accordance with HRTEM analyses, SAED patterns and their azimuthally integrated profiles (Figure S9) for the thick NWs exhibited sharper Bragg peaks, due to larger size of the coherently scattering domain. While 10 nm and 5.1 nm-thick NWs featured the distinctive peaks for the orthorhombic phase, labelled by arrows, the phase of 3.4 nm NWs could not be undoubtedly stated based on SAED pattern taken from a bundle of them, due to their smaller width.

The shape of the NWs was also verified by atomic force microscopy (AFM). Figure 3a shows the typical aspect of a single NW. A profile along the white line traced in Figure 3a is shown in Figure 3b. According to AFM, the shape of the NW was not cylindrical, corroborating what found by HRTEM: the cross section indicated that the single wires presented a planar structure. The 3D representation of the NWs (Figure 3c) evidences that the upper surface of the NWs is flat. The average thickness of the NWs, derived from the analysis of single line profiles obtained from 10 NWs from 3 different samples, was 8.1±0.4 nm (n=36), while the lateral size of the upper face of the NWs was 5.5±0.6 nm (n=36), in agreement with the lateral size determined by size statistics based on BF-TEM images. Note that, except for the flat upper surface, the lateral size and shape of the NWs is affected by tip enlargement effect.[21] Considering the nominal size of the AFM tip used in the experiments, we estimated the size of the lower surface through the formula reported in ref. [22], and obtained the same value as measured for the upper facet. Hence the NWs had rectangular section.

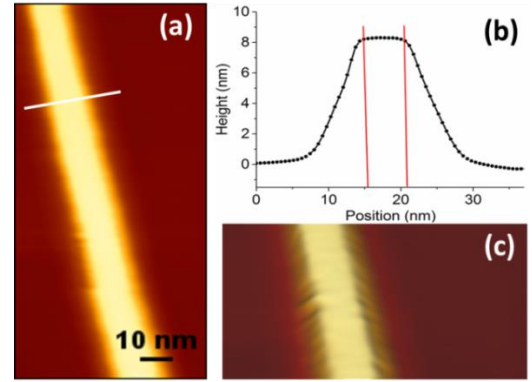

**Figure 3.** (a) Tapping mode AFM height image of a single NW. (b) cross-section along the profile defined by the white line reported in (a). The flat upper part of a NW is clearly visible in the three-dimensional reconstruction of the NW topography reported in panel (c).

**Table 1.** Comparison of the absorption and photoluminescence peak position, PL FWHM, PLQY and PL lifetime for $CsPbBr_3$ NWs with different widths.

| Width (nm) | Abs. max (nm) | PL max (nm) | PL FWHM (nm) | PLQY (%) | Average PL lifetime (ns) |
|---|---|---|---|---|---|
| 20±3 | 511 | 524 | 18 | 12 ± 2 | 20.6 |
| 10±1 | 504 | 517 | 16 | 38 ± 4 | 16.4 |
| 5.1±0.5 | 484 | 496 | 16 | 77 ± 8 | 4.9 |
| 4.1±0.7 | 472 | 481, 491 | 18 | 40 ± 4 | 2.8 |
| 3.4±0.5 | 455, 467 | 473, 483 | 33 | 30 ± 3 | 2.5 |

We recorded UV–visible absorption and PL spectra and measured PLQY and PL lifetime of the various NW samples. The relevant parameters are summarized in Table 1. The presence of strong quantum confinement in the NWs is demonstrated by the optical absorption and PL spectra of Figure 4a (note that the exciton Bohr diameter for $CsPbBr_3$ is around 7 nm[23]): by shrinking the NWs width, the main excitonic absorption peak shifted from 511 nm (for a width $W$ of 10 nm) to 455 nm ($W$ = 3.4 nm), accompanied by a PL peak blue shift from 524 to 473 nm (see Table 1, the full width at half maximum, FWHM, ranged from 16 to 33 nm). Such tunability in the PL peak position is similar to what observed for $CsPbBr_3$ quantum dots,[24] nanoplatelets[20] and hybrid organolead halide NWs[14] exhibiting different degrees of quantum confinement. Here, both NW samples with 4.1±0.7 and 3.4±0.5 nm width evidenced a secondary PL peak (at around 483 and 491 nm, respectively), and a second absorption peak at 467 nm was seen for the $W$ = 3.4 nm sample. These additional spectral features are not seen during the synthesis. In Figure S10 we report for example the PL spectrum from the 4.1±0.7nm NWs directly after the final washing step and immediately after resuspension in toluene. This was characterized by a single emission peak at 473 nm (FWHM 33 nm). However, less than 15 min later, a second PL peak was observed at longer wavelengths. This additional PL peaks is likely due to aggregation of the NWs, which reduces the quantum confinement by allowing delocalization of the holes/electrons on neighbouring NWs. Hence the thinnest NWs had limited stability over time.

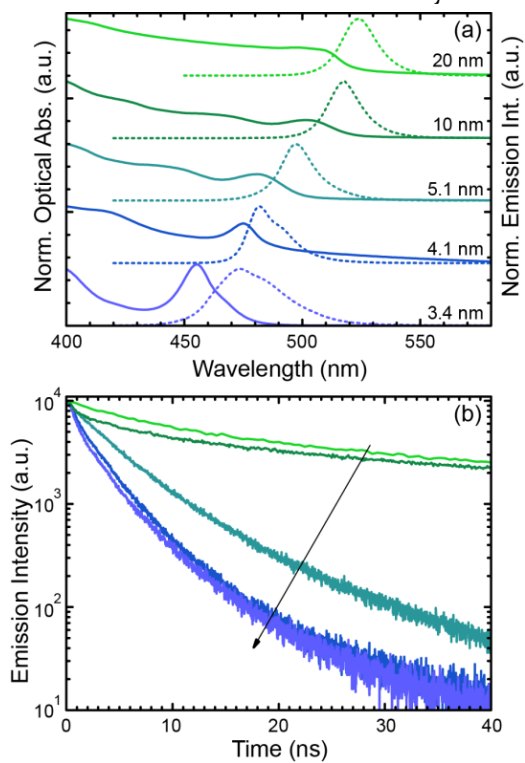

**Figure 4.** (a) Optical absorption (solid line) and PL spectra (dashed line, $\lambda_{exc}$=400 nm) of the NWs in toluene solutions. (b) Time-resolved PL, ($\lambda_{exc}$=405 nm) measured at the PL peak for the $CsPbBr_3$ NWs in toluene solution. Color coding is the same in panels (a) and (b).

The PLQY measured in solution (Table 1) increased from 12% to 77% when the width was reduced from 20±3 to 5.1±0.5 nm. By further reducing the NW width, the PLQY dropped to 30%. The reduction in PLQY was accompanied by the appearance of a second PL peak in the thinnest NWs. This suggests that the red-shifted secondary PL peak has a detrimental effect on the PLQY. Time-resolved PL measurements (Figure 4b) evidenced on the other hand a continuous decrease in PL lifetime, from the green to blue emitting NW samples, from 20.6 ns for 20 nm width to 2.5 ns for 3.4 nm width (Table 1). The NWs with 5.1 nm width, which had the highest PLQY (77%), exhibited a substantially short PL lifetime of 4.9 ns that is in line with that of other CsPbBr3 NCs of different shapes.[20, 25, 26] The increase in PLQY from the thickest (20 nm) to the 5.1 nm NWs, accompanied by a shortening of the PL lifetime, indicates a continuous increase in the radiative rate. Considering that the various samples here reported were prepared under different syntheses conditions, we cannot ascribe the variation in PLQY only to the effect of quantum confinement. It is indeed possible that each sample was characterized by a different number of trap states (hence overall material quality). For the NW samples with width smaller than 5.1 nm (last two rows of Table 1) the decrease in PLQY and the further shortening of the PL lifetime indicate an increase in non-radiative decay rate that can again be tentatively ascribed to the appearance of the secondary PL arising from either NWs aggregation or trap-emission.

In summary, we have reported the colloidal synthesis of $CsPbBr_3$ perovskite NWs with tunable width, from the non-confined regime to strong quantum-confinement regime, by introducing carboxylic acids with short aliphatic chains (octanoic acid or hexanoic acid). The NWs had photoluminescence quantum yield that could be as high as 77%, with PL spectral position that could be varied from green to blue. NWs with a width below ~5 nm show a reduced stability with the appearance of additional PL and absorption peaks and a reduction in PLQY. Future progress in this direction will require the stabilization of the thinnest wires. Another challenge will reside in understanding and modeling the growth kinetics and thermodynamics of these nanostructures. Also, as syntheses protocols to $CsPbBr_3$ perovskite nanostructures of various shapes have reached maturity, interesting developments can be the study of the effect of shape/dimensionality on quantum confinement and on the rate and extent of anion exchange.

## ASSOCIATED CONTENT

### Supporting Information

Experimental details on conditions for the various syntheses, additional data from structural analysis (TEM, HRTEM,

STEM, selected area electron diffraction) and additional data from optical characterization.

## AUTHOR INFORMATION

### Corresponding Author

mirko.prato@iit.it, liberato.manna@iit.it### Notes

The authors declare no competing financial interests.

## ACKNOWLEDGMENT

The research leading to these results has received funding from the European Union 7th Framework Programme under Grant Agreement No. 614897 (ERC Consolidator Grant "TRANS-NANO").

TOC

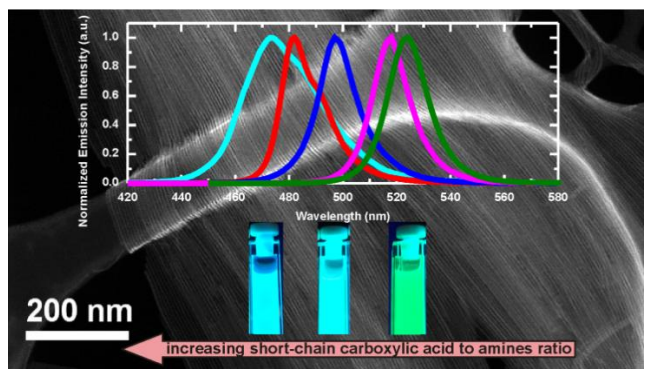